\documentclass[twocolumn,amsmath,amsfont,amssymb,superscriptaddress]{paper}

\usepackage[english]{babel}
\usepackage{graphicx}                       % Include figure files
\usepackage{dcolumn}                        % Align table columns on decimal point
\usepackage[colorlinks=false]{hyperref}     % Hyper references
\usepackage{bm}                             % bold math
\usepackage{subfigure}
\usepackage{longtable}
\usepackage{mathtools}

\usepackage{fullpage}

\usepackage{kantlipsum}
\usepackage{multirow}
\usepackage{longtable}

\usepackage{sidecap}
\usepackage{stmaryrd}
\usepackage{color}
\usepackage{setspace}

\newcommand*{\affaddr}[1]{#1} % No op here. Customize it for different styles.
\newcommand*{\affmark}[1][*]{\textsuperscript{#1}}

\begin{document}

\author{
\large Yannick Leo\affmark[1], M\'arton Karsai\affmark[1,*], Carlos Sarraute\affmark[2] and Eric Fleury\affmark[1] \vspace{.05in}\\
\affaddr{\affmark[\scriptsize 1]\small Univ Lyon, ENS de Lyon, Inria, CNRS, UCB Lyon 1, LIP UMR 5668, IXXI, F-69342, Lyon, France}  \vspace{-.1in}\\
\affaddr{\affmark[\scriptsize 2]\small Grandata Labs, Bartolome Cruz 1818 V. Lopez. Buenos Aires, Argentina}\\
\affaddr{\affmark[\scriptsize *]\small Corresponding author: marton.karsai@ens-lyon.fr}\\
}

\title{Correlations of consumption patterns in social-economic networks}

\date{}
\maketitle{}

\begin{abstract}
We analyze a coupled anonymized dataset collecting the mobile phone communication and bank transactions history of a large number of individuals. After mapping the social structure and introducing indicators of socioeconomic status, demographic features, and purchasing habits of individuals we show that typical consumption patterns are strongly correlated with identified socioeconomic classes leading to patterns of stratification in the social structure. In addition we measure correlations between merchant categories and introduce a correlation network, which emerges with a meaningful community structure. We detect multivariate relations between merchant categories and show correlations in purchasing habits of individuals. Our work provides novel and detailed insight into the relations between social and consuming behaviour with potential applications in recommendation system design.
\end{abstract}


% \fi
%\IEEEpeerreviewmaketitle

\section{Introduction}

The consumption of goods and services is a crucial element of human welfare. The uneven distribution of consumption power among individuals goes hand in hand with the emergence and reservation of socioeconomic inequalities in general. Individual financial capacities restrict personal consumer behaviour, arguably correlate with one's purchasing preferences, and play indisputable roles in determining the socioeconomic position of an ego in the larger society~\cite{Deaton1992Understanding,Deaton1980Economics,Piketti2014Capital,Sernau2012Social,Hurst2015Social}. Investigation of relations between these characters carries a great potential in understanding better rational social-economic behaviour~\cite{Fisher1987Fisher}, and project to direct applications in personal marketing, recommendation, and advertising.

Social Network Analysis (SNA) provides one promising direction to explore such problems ~\cite{Wasserman1994Social}, due to its enormous benefit from the massive flow of human behavioural data provided by the digital data revolution~\cite{Lohr2012The}. The advent of this era was propagated by some new data collection techniques, which allowed the recording of the digital footprints and interaction dynamics of millions of individuals~\cite{lazerscience2009, Abraham2010Computational}. On the other hand, although social behavioural data brought us detailed knowledge about the structure and dynamics of social interactions, it commonly failed to uncover the relationship between social and economic positions of individuals. Nevertheless, such correlations play important roles in determining one's socioeconomic status (SES)~\cite{Bourdieu1984}, social tie formation preferences due to status homophily~\cite{McPherson2001Birds,Lazarsfeld1954Friendship}, and in turn potentially stand behind the emergent stratified structure and segregation on the society level~\cite{Sernau2012Social,Grusky2011Theories}. However until now, the coupled investigation of individual social and economic status remained a great challenge due to lack of appropriate data recording such details simultaneously.

As individual economic status restricts one's capacity in purchasing goods and services, it induces divergent consumption patterns between people at different socioeconomic positions~\cite{Fisher1987Fisher, Deaton1992Understanding,Deaton1980Economics}. This is reflected by sets of commonly purchased products, which are further associated to one's social status~\cite{West2004Conspicuous}. Consumption behaviour has been addressed from various angles considering e.g. environmental effects, socioeconomic position, or social influence coming from connected peers~\cite{Deaton1992Understanding}. However, large data-driven studies combining information about individual purchasing and interaction patterns in a society large population are still rare, although questions about correlations between consumption and social behaviour are of utmost interest.

In this study we address these crucial problems via the analysis of a dataset, which simultaneously records the mobile-phone communication, bank transaction history, and purchase sequences of millions of inhabitants of a single country over several months. This corpus, one among the firsts at this scale and details, allows us to infer the socioeconomic status, consumption habits, and the underlying social structure of millions of connected individuals. Using this information our overall goal is to identify people with certain financial capacities, and to understand \textit{how much money they spend, on what they spend, and whether they spend like their friends?} More precisely, we formulate our study around two research questions:
\begin{itemize}
\item Can one associate typical consumption patterns to people and to their peers belonging to the same or different socioeconomic classes, and if yes how much such patterns vary between individuals or different classes?
\item Can one draw relations between commonly purchased goods or services in order to understand better individual consumption behaviour?
\end{itemize}
After reviewing the related literature in Section~\ref{sec:relatedwork}, we describe our dataset in Section~\ref{sec:data}, and introduce individual socioeconomic indicators to define socioeconomic classes in Section~\ref{sec:sociomeas}. In Section~\ref{sec:purchsocio} we show how typical consumption patterns vary among classes and relate them to structural correlations in the social network. In Section~\ref{sec:purchnet} we draw a correlation network between consumption categories to detect patterns of commonly purchased goods and services. Finally we present some concluding remarks and future research ideas.

\section{Related work}
\label{sec:relatedwork}

Earlier hypothesis on the relation between consumption patterns and socioeconomic inequalities, and their correlations with demographic features such as age, gender, or social status were drawn from specific sociological studies~\cite{chan2010social} and from cross-national social surveys~\cite{deaton1997analysis}. However, recently available large datasets help us to effectively validate and draw new hypotheses as population-large individual level observations and detailed analysis of human behavioural data became possible. These studies shown that personal social interactions, social influence \cite{Deaton1992Understanding}, or homophily~\cite{Wood2012Social} in terms of age or gender~\cite{kovanen2013temporal} have strong effects on purchase behaviour, knowledge which led to the emergent domain of online social marketing~\cite{Felix2016Elements}. Yet it is challenging to measure correlations between individual social status, social network, and purchase patterns simultaneously. Although socioeconomic parameters can be estimated from communication networks~\cite{dong2014inferring} or from external aggregate data~\cite{eagle2010network} usually they do not come together with individual purchase records. In this paper we propose to explore this question through the analysis of a combined dataset proposing simultaneous observations of social structure, economic status and purchase habits of millions of individuals.

\section{Data description}
\label{sec:data}

In the following we are going to introduce two datasets extracted from a corpus combining the mobile phone interactions with purchase history of individuals.

\subsection*{DS1: Ego social-economic data with purchase distributions}
Communication data used in our study records the temporal sequence of 7,945,240,548 call and SMS interactions of 111,719,360 anonymized mobile phone users for $21$ consecutive months. Each call detailed record (CDR) contains the time, unique caller and callee encrypted IDs, the direction (who initiate the call/SMS), and the duration of the interaction. At least one participant of each interaction is a client of a single mobile phone operator, but other mobile phone users who are not clients of the actual provider also appear in the dataset with unique IDs. All unique IDs are anonymized as explained below, thus individual identification of any person is impossible from the data. Using this dataset we constructed a large social network where nodes are users (whether clients or not of the actual provider), while links are drawn between any two users if they interacted (via call or SMS) at least once during the observation period. We filtered out call services, companies, and other non-human actors from the social network by removing all nodes (and connected links) who appeared with either in-degree $k_{in}=0$ or out-degree $k_{out}=0$. We repeated this procedure recursively until we received a network where each user had $k_{in}, k_{out}>0$, i. e. made at least one out-going and received at least one in-coming communication event during the nearly two years of observation. After construction and filtering the network remained with 82,453,814 users connected by 1,002,833,289 links, which were considered to be undirected after this point.

To calculate individual economic estimators we used a dataset provided by a single bank. This data records financial details of 6,002,192 people assigned with unique anonymized identifiers over $8$ consecutive months. The data provides time varying customer variables as the amount of their debit card purchases, their monthly loans, and static user attributes such as their billing postal code (zip code), their age and their gender.

A subset of IDs of the anonymized bank and mobile phone customers were matched\footnote{The matching, data hashing, and anonymization procedure was carried out without the involvement of the scientific partner. After this procedure only anonymized hashed IDs were shared disallowing the direct identification of individuals in any of the datasets.}. This way of combining the datasets allowed us to simultaneously observe the social structure and estimate economic status (for definition see Section~\ref{sec:sociomeas}) of the connected individuals. This combined dataset contained 999,456 IDs, which appeared in both corpuses. However, for the purpose of our study we considered only the largest connected component of this graph. This way we operate with a connected social graph of 992,538 people connected by 1,960,242 links, for all of them with communication events and detailed bank records available.

To study consumption behaviour we used purchase sequences recording the time, amount, merchant category code of each purchase event of each individual during the observation period of $8$ months. Purchase events are linked to one of the 281 merchant category codes (mcc) indicating the type of the actual purchase, like fast food restaurants, airlines, gas stations, etc. Due to the large number of categories in this case we decided to group mccs by their types into $28$ purchase category groups (PCGs) using the categorization proposed in~\cite{MCCAmExp}. After analyzing each purchase groups $11$ of them appeared with extremely low activity representing less than 0.3\% (combined) of the total amount of purchases, thus we decided to remove them from our analysis and use only the remaining $K_{17}$ set of $17$ groups (for a complete list see Fig.\ref{fig:2}a). Note that the group named \emph{Service Providers} ($k_1$ with mcc $24$) plays a particular role as it corresponds to cash retrievals and money transfers and it represents around $70\%$ of the total amount of purchases. As this group dominates over other ones, and since we have no further information how the withdrawn cash was spent, we analyze this group $k_{1}$ separately from the other $K_{2\text{-}17}=K_{17}\backslash\{k_1\}$ set of groups.

This way we obtained DS1, which collects the social ties, economic status, and coarse grained purchase habit informations of $\sim 1$ million people connected together into a large social network.

\subsection*{DS2: Detailed ego purchase distributions with age and gender}

From the same bank transaction trace of 6,002,192 users, we build a second data set DS2. This dataset collects data about the age and gender of individuals together with their purchase sequence recording the time, amount, and mcc of each debit card purchase of each ego. To obtain a set of active users we extracted a corpus of 4,784,745 people that were active at least two months during the observation period. Then for each ego, we assigned a feature set $PV(u):\{ age_u, gender_u, SEG_u, r(c_i,u) \}$ where SEG assigns a socioeconomic group (for definition see Section~\ref{sec:sociomeas}) and $r(c_i,u)$ is an ego purchase distribution vector defined as
\begin{equation}
r(c_i,u)=\frac{m_u^{c_i}}{\sum_{c_i} m_u^{c_i}}.
\end{equation}
This vector assigns the fraction of $m_u^{c_i}$ money spent by user $u$ on a merchant category $c_i$ during the observation period. We excluded purchases corresponding to cash retrievals and money transfers, which would dominate our measures otherwise. A minor fraction of purchases are not linked to valid mccs, thus we excluded them from our calculations.

This way DS2 collects 3,680,652 individuals, without information about their underlying social network, but all assigned with a $PV(u)$ vector describing their personal demographic and purchasing features in details.

\section{Measures of socioeconomic position}
\label{sec:sociomeas}

\begin{figure}[h!]
\centering
\includegraphics[width=0.5\textwidth]{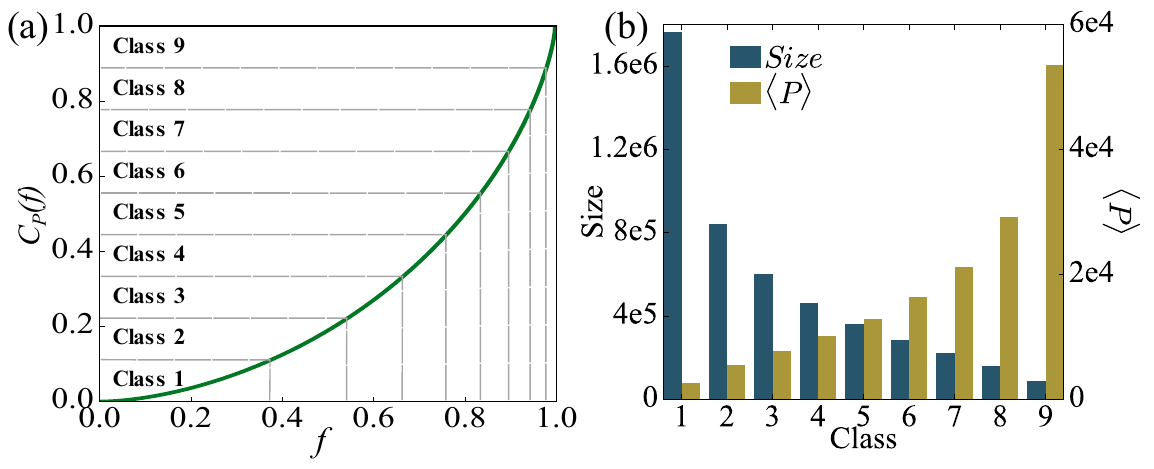}
\caption{\textbf{Social class characteristics} \textbf{(a)} Schematic demonstration of user partitions into 9 socioeconomic classes by using the cumulative AMP function $C(f)$. Fraction of egos belonging to a given class ($x$ axis) have the same sum of AMP $(\sum_u P_u)/n$ ($y$ axis) for each class. \textbf{(b)} Number of egos (green) and the average AMP $\langle P \rangle$ (in USD) per individual (yellow) in different classes.
\label{fig:1}}
\end{figure}

To estimate the personal economic status we used a simple measure reflecting the consumption power of each individual. Starting from the raw data of DS2, which collects the amount and type of debit card purchases, we estimated the economic position of individuals as their average monthly purchase (AMP). More precisely, in case of an ego $u$ who spent $m_u(t)$ amount in month $t$ we calculated the AMP as
\begin{equation}
P_u=\frac{\sum_{t\in T}m_u(t)}{|T|_u}
\end{equation}
where $|T|_u$ corresponds to the number of active months of user $u$ (with at least one purchase in each month). After sorting people by their AMP values we computed the normalized cumulative distribution function of $P_u$ as
\begin{equation}
C(f)=\frac{\sum_{f'=0}^{f} P_u(f')}{\sum_{u} P_u}
\end{equation}
as a function of $f$ fraction of people. This function (Fig.\ref{fig:1}a) appears with high variance and suggests large imbalances in terms of the distribution of economic capacities among individuals in agreement with earlier social theory~\cite{Pareto1971Manual}.

Subsequently we used the $C(f)$ function to assign egos into 9 economic classes (also called socioeconomic classes with smaller numbers assigning lower classes) such that the sum of AMP in each class $s_j$ was the same equal to $(\sum_u P_u)/n$ (Fig.\ref{fig:1}). We decided to use $9$ distinct classes based on the common three-stratum model~\cite{Brown2009Social}, which identifies three main social classes (lower, middle, and upper), and for each of them three sub-classes~\cite{Saunders1990Social}. There are several advantages of this classification: (a) it relies merely on individual economic estimators, $P_u$, (b) naturally partition egos into classes with decreasing sizes for richer groups and (c) increasing $\langle P \rangle$ average AMP values per egos (Fig.\ref{fig:1}b).

\section{Socioeconomic correlations in purchasing patterns}
\label{sec:purchsocio}

In order to address our first research question we were looking for correlations between individuals in different socioeconomic classes in terms of their consumption behaviour on the level of purchase category groups. We analyzed the purchasing behaviour of people in DS1 after categorizing them into socioeconomic classes as explained in Section~\ref{sec:sociomeas}.

\begin{figure*}[ht!]
\centering
\includegraphics[width=0.93\textwidth,angle=0]{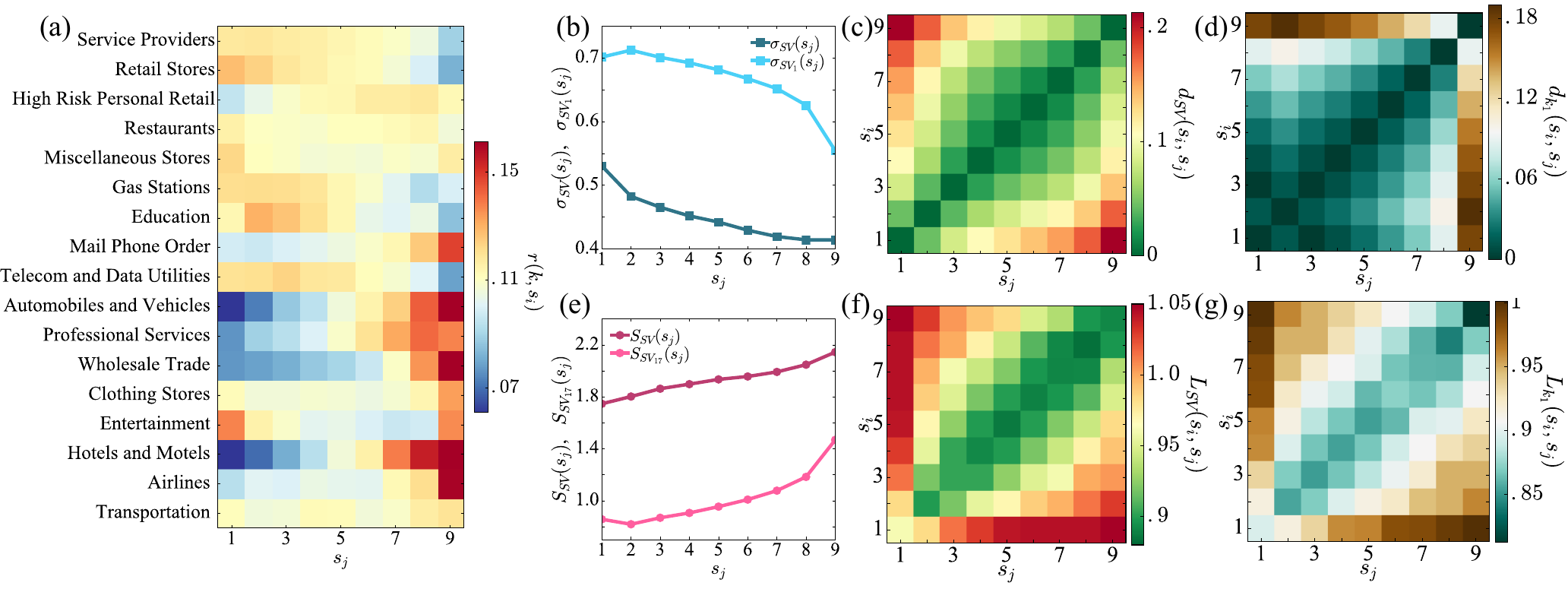}
\caption{\textbf{Consumption correlations in the socioeconomic network} \textbf{(a)} $r(k,s_i)$ distribution of spending in a given purchase category group $k\in K_{17}$ by different classes $s_j$. Distributions are normalised as in Eq.\ref{eq:rks}, i.e. sums up to $1$ for each category. \textbf{(b)} Dispersion $\sigma_{SV}(s_j)$ for different socioeconomic classes considering PCGs in $K_{2\text{-}17}$ (dark blue) and the single category $k_1$ (light blue). \textbf{(c)} (resp. \textbf{(d)}) Heat-map matrix representation of $d_{SV}(s_i,s_j)$ (resp. $d_{k_{1}}(s_i,s_j)$) distances between the average spending vectors of pairs of socioeconomic classes considering PCGs in $K_{2\text{-}17}$ (resp. $k_1$). \textbf{(e)} Shannon entropy measures for different socioeconomic classes considering PCGs in $K_{2\text{-}17}$ (dark pink) and in $k_{17}$ (light pink). \textbf{(f)} (resp. \textbf{(g)}) Heat-map matrix representation of the average $L_{SV}(s_i,s_j)$ (resp. $L_{k_1}(s_i,s_j)$) measure between pairs of socioeconomic classes considering PCGs in $K_{2\text{-}17}$ (resp. $k_1$).
\label{fig:2}}
\end{figure*}

First for each class $s_j$ we take every user $u\in s_j$ and calculate the $m_u^k$ total amount of purchases they spent on a purchase category group $k\in K_{17}$. Then we measure a fractional distribution of spending for each PCGs as:
\begin{equation}
r(k,s_j)=\frac{\sum_{u\in s_j} m^k_u}{\sum_{u\in s} m^k_u},
\label{eq:rks}
\end{equation}
where $s=\bigcup_{j}s_j$ assigns the complete set of users. In Fig.\ref{fig:2}a each line shows the $r(k,s_j)$ distributions for a PCG as the function of $s_j$ social classes, and lines are sorted (from top to bottom) by the total amount of money spent on the actual PCG\footnote{Note that in our social class definition the cumulative AMP is equal for each group and this way each group represents the same economic potential as a whole. Values shown in Fig.\ref{fig:2}a assign the total purchase of classes. Another strategy would be to calculate per capita measures, which in turn would be strongly dominated by values associated to the richest class, hiding any meaningful information about other classes.}. Interestingly, people from lower socioeconomic classes spend more on PCGs associated to essential needs, such as \emph{Retail Stores (St.)}, \emph{Gas Stations},  \emph{Service Providers} (cash) and \emph{Telecom}, while in the contrary, other categories associated to extra needs such as  \emph{High Risk Personal Retail} (Jewelry, Beauty), \emph{Mail Phone Order}, \emph{Automobiles}, \emph{Professional Services (Serv.)} (extra health services), \emph{Whole Trade} (auxiliary goods), \emph{Clothing St.}, \emph{Hotels} and \emph{Airlines} are dominated by people from higher socioeconomic classes. Also note that concerning \emph{Education} most of the money is spent by the lower middle classes, while  \emph{Miscellaneous St.} (gift, merchandise, pet St.) and more apparently \emph{Entertainment} are categories where the lowest and highest classes are spending the most.

From this first analysis we can already identify large differences in the spending behaviour of people from lower and upper classes. To further investigate these dissimilarities on the individual level, we consider the $K_{2\text{-}17}$ category set as defined in section~\ref{sec:data} (category $k_1$ excluded) and build a spending vector $SV(u)=[SV_2(u), ..., SV_{17}(u)]$ for each ego $u$. Here each item $SV_{k}(u)$ assigns the fraction of money $m_u^k/m_u$ that user $u$ spent on a category $k \in K_{2\text{-}17}$ out of his/her $m_u=\sum_{k \in K}{m_u^k}$ total amount of purchases. Using these individual spending vectors we calculate the average spending vector of a given socioeconomic class as $\overline{SV}(s_j) =\langle SV(u) \rangle_{u\in s_j}$. We associate $\overline{SV}(s_j)$ to a representative consumer of class $s_j$ and use this average vector to quantify differences between distinct socioeconomic classes as follows.

The euclidean metric between average spending vectors is:
\begin{equation}
d_{SV}(s_i,s_j) = \lVert \overline{SV}_k(s_i)-\overline{SV}_k(s_j)\rVert_2,
\end{equation}
where $ \lVert \vec{v}  \rVert_2=\sqrt{\sum_k v_k^2}$ assigns the $L^2$ norm of a vector $\vec{v}$. Note that the diagonal elements of $d_{SV}(s_i,s_i)$ are equal to zero by definition. However, in Fig.\ref{fig:2}c the off-diagonal green component around the diagonal indicates that the average spending behaviour of a given class is the most similar to neighboring classes, while dissimilarities increase with the gap between socioeconomic classes. We repeated the same measurement separately for the single category of cash purchases (PCG $k_1$). In this case euclidean distance is defined between average scalar measures as $d_{k_1}(s_i,s_j)=\lVert \langle SV_1\rangle (s_i)-\langle SV_1\rangle(s_j) \rVert_2$. Interestingly, results shown in Fig.\ref{fig:2}d. indicates that here the richest social classes appear with a very different behaviour. This is due to their relative underspending in cash, which can be also concluded from Fig.\ref{fig:2}a (first row). On the other hand as going towards lower classes such differences decrease as cash usage starts to dominate.

To explain better the differences between socioeconomic classes in terms of purchasing patterns, we introduce two additional scalar measures. First, we introduce the dispersion of individual spending vectors as compared to their class average as
\begin{equation}
\sigma_{SV}(s_j) = \langle \lVert \overline{SV}_k(s_j)-SV_k(u) \rVert_2 \rangle_{u\in s_j},
\end{equation}
which appears with larger values if people in a given class allocate their spending very differently. Second, we also calculate the Shannon entropy of spending patterns as
\begin{equation}
S_{SV}(s_j) = \sum_{k\in K_{2\text{-} 17}}{-\overline{SV}_k(s_j) \log(\overline{SV}_k(s_j))}
\end{equation}
to quantify the variability of the average spending vector for each class. This measure is minimal if each ego of a class $s_j$ spends exclusively on the same single PCG, while it is maximal if they equally spend on each PCG. As it is shown in Fig.\ref{fig:2}b (light blue line with square symbols) dispersion decreases rapidly as going towards higher socioeconomic classes. This assigns that richer people tends to be more similar in terms of their purchase behaviour. On the other hand, surprisingly, in Fig.\ref{fig:2}e (dark pink line with square symbols) the increasing trend of the corresponding entropy measure suggests that even richer people behave more similar in terms of spending behaviour they used to allocate their purchases in more PCGs. These trends are consistent even in case of $k_1$ cash purchase category (see $\sigma_{SV_1}(s_j)$ function depicted with dark blue line in in Fig.\ref{fig:2}b) or once we include category $k_1$ into the entropy measure $S_{SV_{17}}(s_j)$ (shown in Fig.\ref{fig:2}b with light pink line).

\begin{figure*}[ht!]
\centering
\includegraphics[width=0.9\textwidth,angle=0]{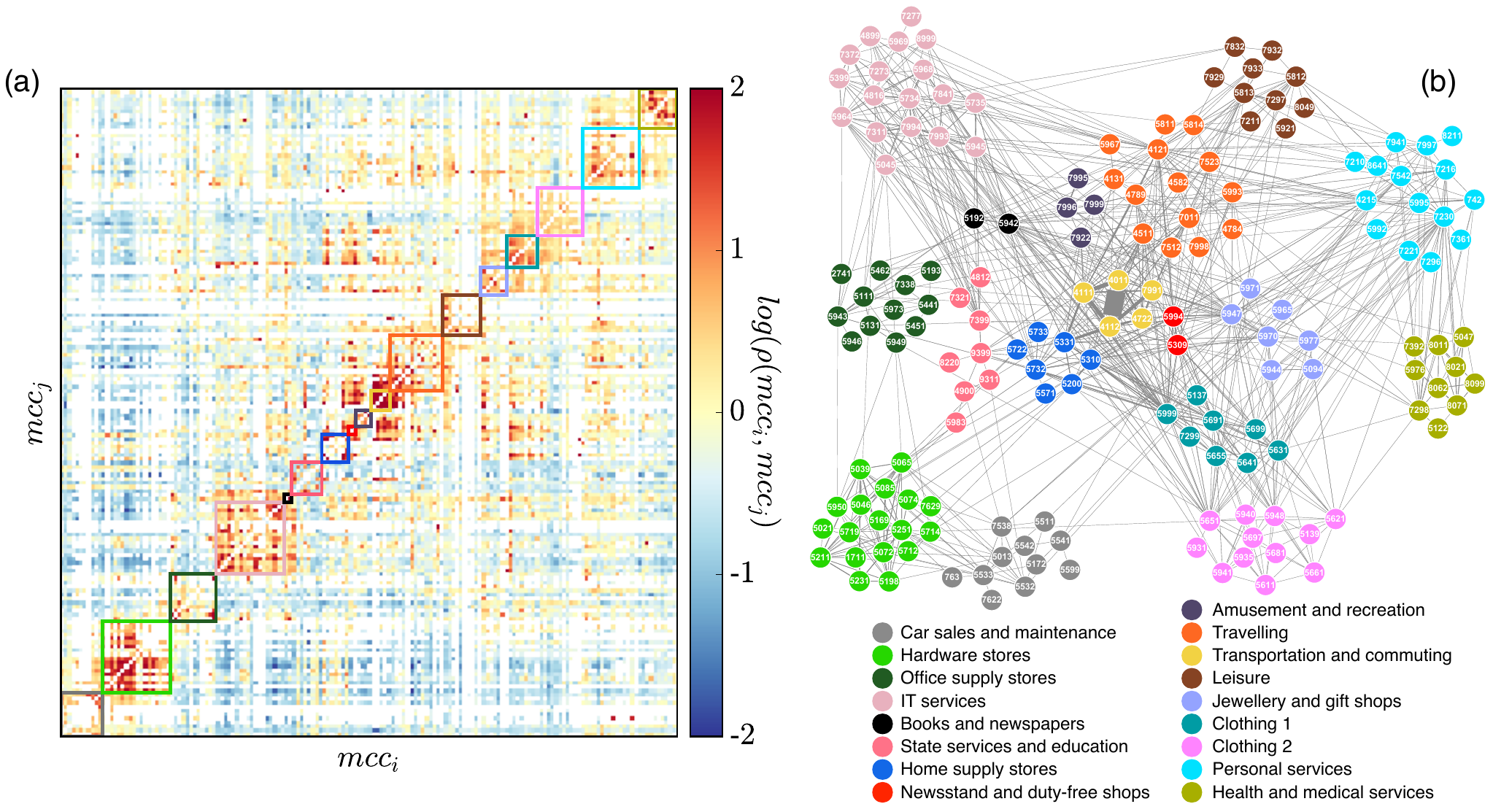}
\caption{\textbf{Merchant category correlation matrix and graph} \textbf{(a)} 163$\times$163 matrix heatmap plot corresponding to $\rho(c_{i}, c_{j})$ correlation values (see Eq.~\ref{eq:corMcc}) between categories. Colors scale with the logarithm of correlation values. Positive (resp. negative) correlations are assigned by red (resp. blue) colors. Diagonal components represent communities with frames colored accordingly.\textbf{(b)} Weighted $G_{\rho}^>$ correlation graph with nodes  annotated with MCCs (see Table~\ref{table:mcc}). Colors assign 17 communities of merchant categories with representative names summarized in the figure legend.}
\label{fig:3}
\end{figure*}

To complete our investigation we characterize the effects of social relationships on the purchase habits of individuals. We address this problem through an overall measure quantifying differences between individual purchase vectors of connected egos positioned in the same or different socioeconomic classes. More precisely, we consider each social tie $(u,v)\in E$ connecting individuals $u\in s_i$ and $v\in s_j$, and for each purchase category $k$ we calculate the average absolute difference of their purchase vector items as
\begin{equation}
d^k(s_i,s_j)=\langle | SV_k(u)-SV_k(v)|\rangle_{u\in s_i, v\in s_j}.
\end{equation}
Following that, as a reference system we generate a corresponding configuration network by taking randomly selected edge pairs from the underlying social structure and swap them without allowing multiple links and self loops. In order to vanish any residual correlations we repeated this procedure in $5\times |E|$ times. This randomization keeps the degree, individual economic estimators $P_u$, the purchase vector $SV(u)$, and the assigned class of each people unchanged, but destroys any structural correlations between egos in the social network, consequently between socioeconomic classes as well. After generating a reference structure we computed an equivalent measure $d_{rn}^k(s_i,s_j)$ but now using links $(u,v)\in E_{rn}$ of the randomized network. We repeated this procedure $100$ times and calculated an average $\langle d_{rn}^k\rangle (s_i,s_j)$. In order to quantify the effect of the social network we simply take the ratio
\begin{equation}
L_k(s_i,s_j)=\frac{d^k(s_i,s_j)}{\langle d_{rn}^k\rangle (s_i,s_j)}
\end{equation}
and calculate its average $L_{SV}(s_i,s_j)=\langle L_k(s_i,s_j)\rangle_k$ over each category group $k\in K_{2\text{-}17}$ or respectively $k_1$. This measure shows whether connected people have more similar purchasing patterns than one would expect by chance without considering any effect of homophily, social influence or structural correlations. Results depicted in Fig.\ref{fig:2}f and \ref{fig:2}g for $L_{SV}(s_i,s_j)$ (and $L_{k_1}(s_i,s_j)$ respectively) indicates that the purchasing patterns of individuals connected in the original structure are actually more similar than expected by chance (diagonal component). On the other hand people from remote socioeconomic classes appear to be less similar than one would expect from the uncorrelated case (indicated by the $L_{SV}(s_i,s_j)>1$ values typical for upper classes in Fig.\ref{fig:2}f). Note that we found the same correlation trends in cash purchase patterns as shown in Fig.\ref{fig:2}g. These observations do not clearly assign whether homophily~\cite{McPherson2001Birds,Lazarsfeld1954Friendship} or social influence~\cite{Deaton1992Understanding} induce the observed similarities in purchasing habits but undoubtedly clarifies that social ties (i.e. the neighbors of an ego) and socioeconomic status play deterministic roles in the emerging similarities in consumption behaviour.

\section{Purchase category correlations}
\label{sec:purchnet}

To study consumption patterns of single purchase categories PCGs provides a too coarse grained level of description. Hence, to address our second question we use DS2 and we downscale from the category group level to the level of single merchant categories. We are dealing with 271 categories after excluding some with less than 100 purchases and the categories linked to money transfer and cash retrieval (for a complete list of IDs and name of the purchase categories considered see Table~\ref{table:mcc}). As in Section~\ref{sec:data} we assign to each ego $u$ a personal vector $PV(u)$ of four socioeconomic features: the age, the gender, the social economic group, and the distribution $r(c_i,u)$ of purchases in different merchant categories made by the central ego. Our aim here is to obtain an overall picture of the consumption structure at the level of merchant categories and to understand precisely how personal and socioeconomic features correlate with the spending behaviour of individuals and with the overall consumption structure. 

As we noted in section~\ref{sec:purchsocio}, the purchase spending vector $r(c_{i},u)$ of an ego quantifies the fraction of money spent on a category $c_i$. Using the spending vectors of $n$ number of individuals we define an overall correlation measure between categories as
\begin{equation}
\rho(c_{i}, c_{j})=\frac{n (\sum_{u}{r(c_{i},u) r(c_{j},u)})}{(\sum_{u}{r(c_{i},u)}) (\sum_{u}{r(c_{j},u)})}.
\label{eq:corMcc}
\end{equation}

\begin{figure*}[ht!]
\centering
\includegraphics[width=0.9\textwidth,angle=0]{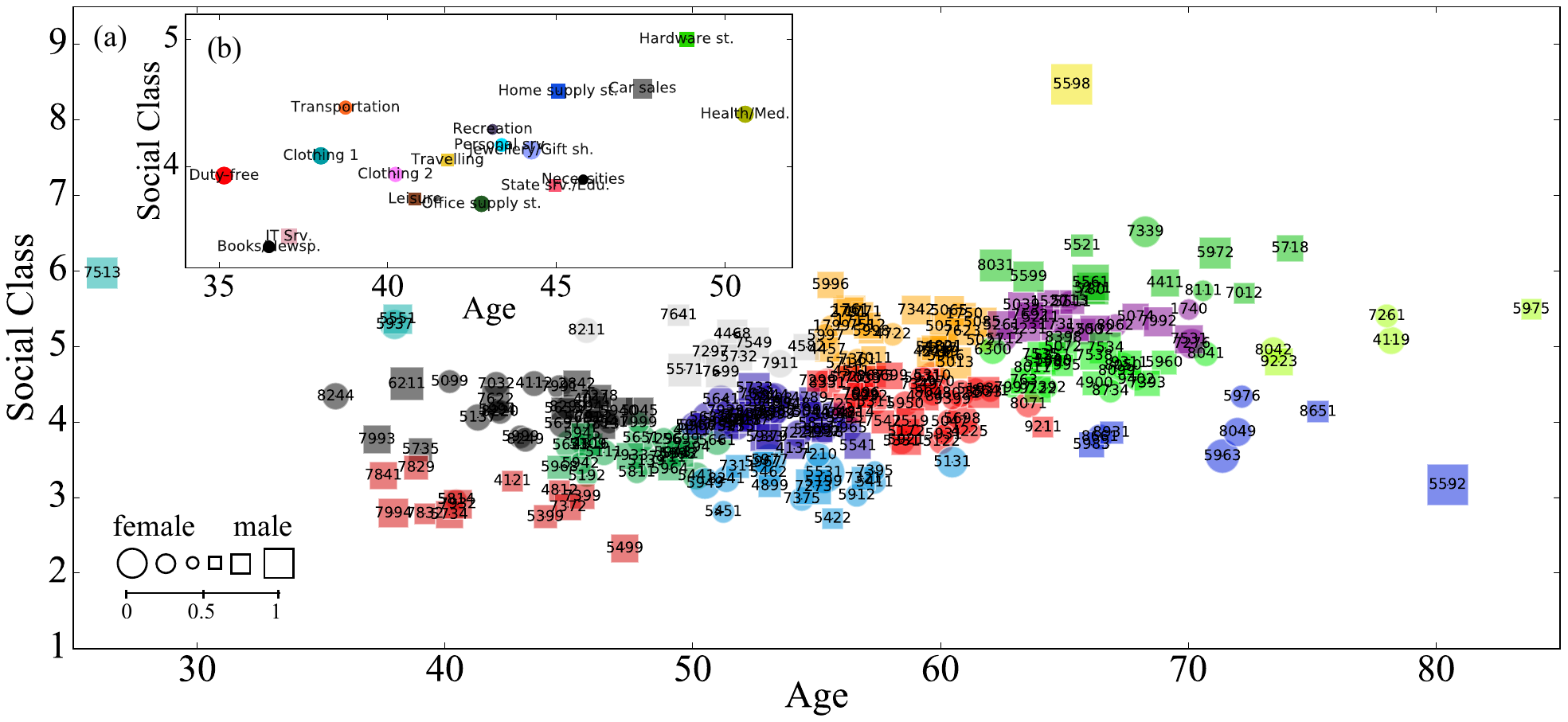}
\caption{\textbf{Socioeconomic parameters of merchant categories} \textbf{(a)} Scatter plot of $AFS(c_i)$ triplets (for definition see Eq.~\ref{eq:vc} and text) for 271 merchant categories summarized in Table~\ref{table:mcc}. Axis assign average age and SEG of purchase categories, while gender information are assigned by symbols. The shape of symbols assigns the dominant gender (circle-female, square-male) and their size scales with average values. \textbf{(b)} Similar scatter plot computed for communities presented in Fig.\ref{fig:3}b. Labels and colors are explained in the legend of Fig.\ref{fig:3}a.\label{fig:4}}
\end{figure*}

This symmetric formulae quantifies how much people spend on a category $c_{i}$ if they spend on an other $c_{j}$ category or vice versa. Therefore, if $\rho(c_{i}, c_{j})>1$, the categories $c_{i}$ and $c_{j}$ are positively correlated and if $\rho(c_{i}, c_{j})<1$, categories are negatively correlated. Using $\rho(c_{i}, c_{j})$ we can define a weighted correlation graph $G_{\rho}=(V_{\rho},E_{\rho},\rho)$ between categories $c_i\in V_{\rho}$, where links $(c_{i}, c_{j})\in E_{\rho}$ are weighted by the $\rho(c_{i}, c_{j})$ correlation values. The weighted adjacency matrix of $G_{\rho}$ is shown in Fig.\ref{fig:3}a as a heat-map matrix with logarithmically scaling colors. Importantly, this matrix emerges with several block diagonal components suggesting present communities of strongly correlated categories in the graph.

To identify categories which were commonly purchased together we consider only links with positive correlations. Furthermore, to avoid false positive correlations, we consider a $10\%$ error on $r$ that can induce, in the worst case $50\%$ overestimation of the correlation values. In addition, to consider only representative correlations we take into account category pairs which were commonly purchased by at least $1000$ consumers. This way we receive a $G_{\rho}^>$ weighted sub-graph of $G_{\rho}$, shown in Fig.\ref{fig:3}b, with 163 nodes and 1664 edges with weights $\rho(c_{i}, c_{j})>1.5$.

To identify communities in $G_{\rho}^>$ indicated by the correlation matrix in Fig.\ref{fig:3}a we applied a graph partitioning method based on the Louvain algorithm~\cite{blondel2008fast}. We obtained 17 communities depicted with different colors in Fig.\ref{fig:3}b and as corresponding colored frames in Fig.\ref{fig:3}a. Interestingly, each of these communities group a homogeneous set of merchant categories, which could be assigned to similar types of purchasing activities (see legend of Fig.\ref{fig:3}b). In addition, this graph indicates how different communities are connected together. Some of them, like \textit{Transportation, IT} or \textit{Personal Serv.} playing a central role as connected to many other communities, while other components like \textit{Car sales and maintenance} and \textit{Hardware St.}, or \textit{Personal} and \textit{Health and medical Serv.} are more like pairwise connected. Some groups emerge as standalone communities like \textit{Office Supp. St.}, while others like \textit{Books and newspapers} or \textit{Newsstands and duty-free Shops (Sh.)} appear as bridges despite their small sizes.

Note that the main categories corresponding to everyday necessities related to food (\emph{Supermarkets}, \emph{Food St.}) and telecommunication (\emph{Telecommunication Serv.}) do not appear in this graph. Since they are responsible for the majority of total spending, they are purchased necessarily by everyone without obviously enhancing the purchase in other categories, thus they do not appear with strong correlations.

Finally we turn to study possible correlations between purchase categories and personal features. An average feature set $AFS(c_i)=\{\langle age(c_i) \rangle,\langle gender(c_i) \rangle, \langle SEG(c_i\} \rangle)$ is assigned to each of the 271 categories. The average $\langle v(c_i) \rangle$ of a feature $v\in \{ age, gender, SEG\}$ assigns a weighted average value computed as:
\begin{equation}
\langle v(c_i) \rangle = \frac{\sum_{u \in \{ u\}_i}\alpha_i(v_u)v_u}{\sum_{u \in \{ u\}_u}\alpha_i(v)},
\label{eq:vc}
\end{equation}
where $v_u$ denotes a feature of a user $u$ from the $\{ u\}_i$ set of individuals who spent on category $c_i$. Here
\begin{equation}
\alpha_i(v_u)=\sum_{(u \in \{ u\}_i | v_u=v)} \frac{r(c_i,u)}{n_i(v_u)}
\end{equation}
corresponds to the average spending on category $c_i$ of the set of users from $\{ u\}_i$ sharing the same value of the feature $v$. $n_i(v_u)$ denotes the number of such users. In other words, e.g. in case of $v=age$ and $c_{742}$, $\langle age(c_{742}) \rangle$ assigns the average age of people spent on Veterinary Services ($mcc=742$) weighted by the amount they spent on it. In case of $v=gender$ we assigned $0$ to females and $1$ to males, thus the average gender of a category can take any real value between $[0, 1]$, indicating more females if $\langle gender(c_i) \rangle\leq 0.5$ or more males otherwise.

\begin{table*}[ht!]
\resizebox{2.1\columnwidth}{!}{%
%\fontsize{6}{7}\selectfont
\centering

  \begin{tabular}{ | @{\hspace{1pt}} l @{\hspace{2pt}} | @{\hspace{1pt}} l @{\hspace{2pt}} | @{\hspace{1pt}} l @{\hspace{2pt}} | @{\hspace{1pt}} l @{\hspace{2pt}} | @{\hspace{1pt}} l @{\hspace{2pt}} | @{\hspace{1pt}} l @{\hspace{2pt}} |}
    \hline
742: Veterinary Serv. & 5072: Hardware Supp. & 5598: Snowmobile Dealers & 5950: Glassware, Crystal St. & 7296: Clothing Rental & 7941: Sports Clubs \\ \hline
763: Agricultural Cooperative & 5074: Plumbing, Heating Equip. & 5599: Auto Dealers & 5960: Dir Mark - Insurance & 7297: Massage Parlors & 7991: Tourist Attractions \\ \hline
780: Landscaping Serv. & 5085: Industrial Supplies & 5611: Men Cloth. St. & 5962: Direct Marketing - Travel & 7298: Health and Beauty Spas & 7992: Golf Courses \\ \hline
1520: General Contr. & 5094: Precious Objects/Stones & 5621: Wom Cloth. St. & 5963: Door-To-Door Sales & 7299: General Serv. & 7993: Video Game Supp. \\ \hline
1711: Heating, Plumbing & 5099: Durable Goods  & 5631: Women?s Accessory Sh. & 5964: Dir. Mark. Catalog & 7311: Advertising Serv. & 7994: Video Game Arcades \\ \hline
1731: Electrical Contr. & 5111: Printing, Office Supp. & 5641: Children?s Wear St. & 5965: Dir. Mark. Retail Merchant & 7321: Credit Reporting Agencies & 7995: Gambling \\ \hline
1740: Masonry \& Stonework & 5122: Drug Proprietaries & 5651: Family Cloth. St. & 5966: Dir Mark - TV & 7333: Graphic Design & 7996: Amusement Parks \\ \hline
1750: Carpentry Contr. & 5131: Notions Goods & 5655: Sports \& Riding St. & 5967: Dir. Mark. & 7338: Quick Copy & 7997: Country Clubs \\ \hline
1761: Sheet Metal & 5137: Uniforms Clothing & 5661: Shoe St. & 5968: Dir. Mark. Subscription & 7339: Secretarial Support Serv. & 7998: Aquariums \\ \hline
1771: Concrete Work Contr. & 5139: Commercial Footwear & 5681: Furriers Sh. & 5969: Dir. Mark. Other & 7342: Exterminating Services & 7999: Recreation Serv. \\ \hline
1799: Special Trade Contr. & 5169: Chemicals Products  & 5691: Cloth. Stores & 5970: Artist?s Supp. & 7349: Cleaning and Maintenance & 8011: Doctors \\ \hline
2741: Publishing and Printing & 5172: Petroleum Products & 5697: Tailors & 5971: Art Dealers \& Galleries & 7361: Employment Agencies & 8021: Dentists, Orthodontists \\ \hline
2791: Typesetting Serv. & 5192: Newspapers & 5698: Wig and Toupee St. & 5972: Stamp and Coin St. & 7372: Computer Programming & 8031: Osteopaths \\ \hline
2842: Specialty Cleaning & 5193: Nursery \& Flowers Supp. & 5699: Apparel Accessory Sh. & 5973: Religious St. & 7375: Information Retrieval Serv. & 8041: Chiropractors \\ \hline
4011: Railroads & 5198: Paints & 5712: Furniture & 5975: Hearing Aids & 7379: Computer Repair & 8042: Optometrists \\ \hline
4111: Ferries & 5199: Nondurable Goods & 5713: Floor Covering St. & 5976: Orthopedic Goods & 7392: Consulting, Public Relations & 8043: Opticians \\ \hline
4112: Passenger Railways & 5200: Home Supply St. & 5714: Window Covering St. & 5977: Cosmetic St. & 7393: Detective Agencies & 8049: Chiropodists, Podiatrists \\ \hline
4119: Ambulance Serv. & 5211: Materials St. & 5718: Fire Accessories St. & 5978: Typewriter St. & 7394: Equipment Rental & 8050: Nursing/Personal Care \\ \hline
4121: Taxicabs & 5231: Glass \& Paint St. & 5719: Home Furnishing St. & 5983: Fuel Dealers (Non Auto) & 7395: Photo Developing & 8062: Hospitals \\ \hline
4131: Bus Lines & 5251: Hardware St. & 5722: House St. & 5992: Florists & 7399: Business Serv. & 8071: Medical Labs \\ \hline
4214: Motor Freight Carriers & 5261: Nurseries \& Garden St. & 5732: Elec. St. & 5993: Cigar St. & 7512: Car Rental Agencies & 8099: Medical Services \\ \hline
4215: Courier Serv. & 5271: Mobile Home Dealers & 5733: Music Intruments St. & 5994: Newsstands & 7513: Truck/Trailer Rentals & 8111: Legal Services, Attorneys \\ \hline
4225: Public Storage & 5300: Wholesale & 5734: Comp.Soft. St. & 5995: Pet Sh. & 7519: Mobile Home Rentals & 8211: Elem. Schools \\ \hline
4411: Cruise Lines & 5309: Duty Free St. & 5735: Record Stores & 5996: Swimming Pools Sales & 7523: Parking Lots, Garages & 8220: Colleges Univ. \\ \hline
4457: Boat Rentals and Leases & 5310: Discount Stores & 5811: Caterers & 5997: Electric Razor St. & 7531: Auto Body Repair Sh. & 8241: Correspondence Schools \\ \hline
4468: Marinas Serv. and Supp. & 5311: Dep. St. & 5812: Restaurants & 5998: Tent and Awning Sh. & 7534: Tire Retreading \& Repair & 8244: Business Schools \\ \hline
4511: Airlines & 5331: Variety Stores & 5813: Drinking Pl. & 5999: Specialty Retail & 7535: Auto Paint Sh. & 8249: Training Schools \\ \hline
4582: Airports, Flying Fields & 5399: General Merch. & 5814: Fast Foods & 6211: Security Brokers & 7538: Auto Service Shops & 8299: Educational Serv. \\ \hline
4722: Travel Agencies & 5411: Supermarkets & 5912: Drug St. & 6300: Insurance & 7542: Car Washes & 8351: Child Care Serv. \\ \hline
4784: Tolls/Bridge Fees & 5422: Meat Prov. & 5921: Alcohol St. & 7011: Hotels & 7549: Towing Serv. & 8398: Donation \\ \hline
4789: Transportation Serv. & 5441: Candy St. & 5931: Secondhand Stores & 7012: Timeshares & 7622: Electronics Repair Sh. & 8641: Associations \\ \hline
4812: Phone St. & 5451: Dairy Products St. & 5932: Antique Sh. & 7032: Sporting Camps & 7623: Refrigeration Repair & 8651: Political Org. \\ \hline
4814: Telecom. & 5462: Bakeries & 5933: Pawn Shops & 7033: Trailer Parks, Camps & 7629: Small Appliance Repair & 8661: Religious Orga. \\ \hline
4816: Comp. Net. Serv. & 5499: Food St. & 5935: Wrecking Yards & 7210: Laundry, Cleaning Serv. & 7631: Watch/Jewelry Repair & 8675: Automobile Associations \\ \hline
4821: Telegraph Serv. & 5511: Cars Sales & 5937: Antique Reproductions & 7211: Laundries & 7641: Furniture Repair & 8699: Membership Org. \\ \hline
4899: Techno St. & 5521: Car Repairs Sales & 5940: Bicycle Sh. & 7216: Dry Cleaners & 7692: Welding Repair & 8734: Testing Lab. \\ \hline
4900: Utilities & 5531: Auto and Home Supp. St. & 5941: Sporting St. & 7217: Upholstery Cleaning & 7699: Repair Sh. & 8911: Architectural Serv. \\ \hline
5013: Motor Vehicle Supp. & 5532: Auto St. & 5942: Book St. & 7221: Photographic Studios & 7829: Picture/Video Production & 8931: Accounting Serv. \\ \hline
5021: Commercial Furniture & 5533: Auto Access. & 5943: Stationery St. & 7230: Beauty Sh. & 7832: Cinema & 8999: Professional Serv. \\ \hline
5039: Constr. Materials & 5541: Gas Stations & 5944: Jewelry St. & 7251: Shoe Repair/Hat Cleaning & 7841: Video Tape Rental St. & 9211: Courts of Law \\ \hline
5044: Photographic Equip. & 5542: Automated Fuel Dispensers & 5945: Toy,-Game Sh. & 7261: Funeral Serv. & 7911: Dance Hall \& Studios & 9222: Government Fees \\ \hline
5045: Computer St. & 5551: Boat Dealers & 5946: Camera and Photo St. & 7273: Dating/Escort Serv. & 7922: Theater Ticket & 9223: Bail and Bond Payments \\ \hline
5046: Commercial Equipment & 5561: Motorcycle Sh. & 5947: Gift Sh. & 7276: Tax Preparation Serv. & 7929: Bands, Orchestras & 9311: Tax Payments \\ \hline
5047: Medical Equipment & 5571: Motorcycle Sh. & 5948: Luggage \& Leather St. & 7277: Counseling Services & 7932: Billiard/Pool & 9399: Government Serv. \\ \hline
5051: Metal Service Centers & 5592: Motor Homes Dealers & 5949: Fabric St. & 7278: Buying/Shopping Serv. & 7933: Bowling & 9402: Postal Serv. \\ \hline
5065: Electrical St. & & & & & \\ \hline
  \end{tabular}
  }
 \vspace{10pt}
\caption{Codes and names of $271$ merchant categories used in our study. MCCs were taken from the Merchant Category Codes and Groups Directory published by American Express~\cite{MCCAmExp}. Abbreviations correspond to: Serv. - Services, Contr. - Contractors, Supp. - Supplies, St. - Stores, Equip. - Equipment, Merch. - Merchandise, Prov. - Provisioners, Pl. - Places, Sh. - Shops, Mark. - Marketing, Univ. - Universities, Org. - Organizations, Lab. - Laboratories.}
\label{table:mcc}
\end{table*}

We visualize this multi-modal data in Fig.\ref{fig:4}a as a scatter plot, where axes scale with average age and SEG, while the shape and size of symbols correspond to the average gender of each category. To further identify correlations we applied k-means clustering~\cite{Bishop1995Neural} using the $AFS(c_i)$ of each category. The ideal number of clusters was $15$ according to several criteria: Davies-Bouldin Criterion, Calinski-Harabasz criterion (variance ratio criterion) and the Gap method \cite{Tibshirani2001Estimating}. Colors in Fig.\ref{fig:4}a assign the identified k-mean clusters.

The first thing to remark in Fig.\ref{fig:4}a is that the average age and SEG assigned to merchant categories are positively correlated with a Pearson correlation coefficient $0.42$ ($p<0.01$). In other words, elderly people used to purchase from more expensive categories, or alternatively, wealthier people tend to be older, in accordance with our intuition. At the same time, some signs of gender imbalances can be also concluded from this plot. Wealthier people appear to be commonly males rather than females. A Pearson correlation measure between gender and SEG, which appears with a coefficient $0.29$ ($p<0.01)$ confirmed it. On the other hand, no strong correlation was observed between age and gender from this analysis.

To have an intuitive insight about the distribution of merchant categories, we take a closer look at specific category codes (summarized in Table~\ref{table:mcc}). As seen in Fig.\ref{fig:4}a elderly people tend to purchase in specific categories such as \emph{Medical Serv.}, \emph{Funeral Serv.}, \emph{Religious Organisations}, \emph{Motorhomes Dealers}, \emph{Donation}, \emph{Legal Serv.}. Whereas categories such as \emph{Fast Foods}, \emph{Video Game Arcades}, \emph{Cinema}, \emph{Record St.}, \emph{Educational Serv.}, \emph{Uniforms Clothing}, \emph{Passenger Railways}, \emph{Colleges-Universities} are associated to younger individuals on average. At the same time, wealthier people purchase more in categories as \emph{Snowmobile Dealers}, \emph{Secretarial Serv.}, \emph{Swimming Pools Sales}, \emph{Car Dealers Sales}, while poorer people tend to purchase more in categories related to everyday necessities like \emph{Food St.}, \emph{General Merch.}, \emph{Dairy Products St.}, \emph{Fast Foods} and \emph{Phone St.}, or to entertainment as \emph{Billiard} or \emph{Video Game Arcades}. Typical purchase categories are also strongly correlated with gender as categories more associated to females are like \emph{Beauty Sh.}, \emph{Cosmetic St.}, \emph{Health and Beauty Spas}, \emph{Women Clothing St.} and \emph{Child Care Serv.}, while others are preferred by males like \emph{Motor Homes Dealers}, \emph{Snowmobile Dealers}, \emph{Dating/Escort Serv.}, \emph{Osteopaths}, \emph{Instruments St.}, \emph{Electrical St.}, \emph{Alcohol St.} and \emph{Video Game Arcades}.

Finally we repeated a similar analysis on communities shown in Fig.\ref{fig:3}b, but computing the $AFS$ on a set of categories that belong to the same community. Results in Fig.\ref{fig:4}b disclose positive age-SEG correlations as observed in Fig.\ref{fig:4}a, together with somewhat intuitive distribution of the communities.

\section{Conclusion}
\label{sec:concl}

In this paper we analyzed a multi-modal dataset collecting the mobile phone communication and bank transactions of a large number of individuals living in a single country. This corpus allowed for an innovative global analysis both in term of social network and its relation to the economical status and merchant habits of individuals. We introduced several measures to estimate the socioeconomic status of each individual together with their purchasing habits. Using these information we identified distinct socioeconomic classes, which reflected strongly imbalanced distribution of purchasing power in the population. After mapping the social network of egos from mobile phone interactions, we showed that typical consumption patterns are strongly correlated with the socioeconomic classes and the social network behind. We observed these correlations on the individual and social class level.

In the second half of our study we detected correlations between merchant categories commonly purchased together and introduced a correlation network which in turn emerged with communities grouping homogeneous sets of categories. We further analyzed some multivariate relations between merchant categories and average demographic and socioeconomic features, and found meaningful patterns of correlations giving insights into correlations in purchasing habits of individuals.

We identified several new directions to explore in the future. One possible track would be to better understand the role of the social structure and interpersonal influence on individual purchasing habits, while the exploration of correlated patterns between commonly purchased brands assigns another promising directions. Beyond our general goal to better understand the relation between social and consuming behaviour these results may enhance applications to better design marketing, advertising, and recommendation strategies, as they assign relations between co-purchased product categories.

\section*{Acknowledgment}
We thank M. Fixman for assistance. We acknowledge the support from the SticAmSud UCOOL project, INRIA, and the SoSweet (ANR-15-CE38-0011-01) and CODDDE (ANR-13-CORD-0017-01) ANR projects. 



%\bibliographystyle{IEEEtran}
%\begin{thebibliography}{1}


%%%%% CLEAR DOUBLE PAGE!
\newpage{\pagestyle{empty}\cleardoublepage}

%\end{thebibliography}

\end{document}